\documentclass[prb,twocolumn,showpacs]{revtex4}
\usepackage{graphicx}
\usepackage{amsfonts}
\usepackage{amsmath}
\usepackage{amssymb}
\usepackage{bm}
\usepackage{hyperref}
\usepackage{slashed}
\begin{document}
\title{Quantum anomalies in superconducting Weyl metals}
\author{Rui Wang$^{1,2}$, Lei Hao$^{1,3}$, Baigeng Wang$^{2}$ C. S. Ting$^{1}$}
\affiliation{$^1$Department of Physics and Texas Center for Superconductivity, University of Houston, Houston, Texas 77204, USA\\
$^2$National Laboratory of Solid State Microstructures and Department of
Physics, Nanjing University, Nanjing 210093, China\\
$^3$Department of Physics, Southeast University, Nanjing 210096, China}
\date{\today }

\begin{abstract}
We theoretically study the quantum anomalies in the superconducting Weyl metals based on the topological field theory. It is demonstrated that the Fermi arc and the surface Andreev bound state, characteristic of the superconducting Weyl metals, are the manifestations of two underlying phenomenon, namely the chiral anomaly and the parity-like anomaly, respectively. The first anomaly is inherited from the Berry curvature around the original Weyl points, while the second is the result of the superconductivity. We show that, all the fascinating topological behavior of the superconducting Weyl metals, either intranode FFLO or the internode BCS pairing state, can be satisfactorily described and predicted by our topological field theory.
\end{abstract}

\pacs{74.20.Mn, 14.80.Va, 03.65.Vf}
\maketitle

\section{introduction}
The topological states of matters have received enormous attention after the discovery \cite{Bernevig,hzhang} and realization \cite{ylchen} of the topological insulators (TIs). TIs is topological in the sense that even though they are fully gapped in the bulk, they enjoy symmetry-protected surface metallic states \cite{Bernevig}. Of particular recent interest is another type of topological phase termed the Weyl semimetals (WSMs) \cite{Wan}, which has no bulk gap but enjoys gapless nodes distributed in its three dimensional (3D) momentum space. It is believed that the linearly dispersed-3D Weyl cone is robust to perturbations and can lead to various of topologically-nontrivial behaviors, including the chiral surface states, \textit{i.e.}, the Fermi arcs terminating at Weyl points with opposite chirality, and the remarkable electromagnetic properties such as the semi-quantized anomalous Hall effect (AHE) \cite{Burkov,yang} and the chiral magnetic effect (CME) \cite{Fukushima}. These properties make it a promising candidate for future application in quantum transport, and therefore stimulates a broad inquiry into this gapless topological phase, both in theory and in experiment.

Theoretically, to reveal the topological aspects, the topological field theory (TFT) can be used to describe the low-energy, universal physics of the topologically ordered states. For example, this method has been applied to TIs \cite{Qi}, which successfully accounted for the AHE and the topological magnetoelectric effect (TME). Besides, it has also been studied in time-reversal invariant topological superconductors, leading to the prediction of the level crossing induced by the crossing of vortex lines \cite{Qib}. Also, the application of the TFT to the WSM phase \cite{Zyuzin,Goswami} brought light to the chiral anomaly and gave rise to the prediction of the AHE and CME. Even though whether the CME persists in realistic materials is still under debating \cite{Vazifeh,Chang,Takane}, the TFT serves as a faithful description of the physically measurable topological response functions. Experimentally, based on recent first principle calculation \cite{Weng,Huang}, the non-centrosymmetric transition metal monophosphides such as TaAs, NbAs, TaP and NbP have been reported \cite{Xu,Xub,Xuc,Lv,nxu,zwang}, showing clear signature of the WSM phase.

Despite the above progress on the WSM phase, recent studies reveal even more interesting perspectives. Remarkably, the first principle calculation \cite{ysun} predicts that $\mathrm{MoTe}_2$ is a WSM state which exhibits four pairs of Weyl points in the bulk band structure. Moreover, $\mathrm{MoTe}_2$ is also  experimentally found to be a superconductor (SC) with transition temperature $T_c=8.2K$ under pressure \cite{yqi}. The above calculation and experiment naturally raises the question: will any new topological properties emerge if a WSM phase becomes a superconductor? Since the superconductor (SC) is based on pairing of electron which usually depends on the density of states (DOS). For undoped WSM, the DOS is zero, which will hamper the SC. Therefore, in order to investigate the SC, the WSM needs to be doped. However, in order to preserve the spin-momentum-locking of the Weyl points (WPs), only a slight doping is allowed, where spherical of Fermi surfaces around the WPs should occur.

The topic of superconductivity in doped WSM (i.e., the superconducting Weyl metal (SWM)) has been investigated in some literatures. Ref.\cite{tmeng}, investigates a superlattice with the staking of layers of TIs and standard s-wave BCS SCs, where it is found that the SC splits one WP into two Bogoliubov-Weyl nodes.  Ref.\cite{Cho,Bednik,Zhou} studied and compared the stability of the internode Bardeen-Cooper-Schrieffer (BCS) pairing state and the intranode Fulde-Ferrell-Larkin-Ovchinnikov (FFLO) pairing state based on the mean-field theory. Moreover, Ref. \cite{Bednik}, Ref.\cite{Zhou} and Ref.\cite{yli} also evaluated the surface states of the SWM phases. It is found that, in the BCS state, both the surface Andreev bound state (SABS) and the Fermi arc will occur on the system's boundary, with the former inside the pairing gap while the latter inside the band gap \cite{Bednik,yli}. In the FFLO state \cite{Zhou}, a pair of Andreev bound states with opposite dispersion slope is extracted, where one locates in the pairing gap around the left-handed Weyl node and the other one resides around the right-handed node. Another pair of SABSs also occurs due to the particle-hole symmetry. Besides, the zero energy excitation in the surface spectrum was found to be the Majorana fermion that is localized in the system's boundary \cite{Zhou,Bednik}. Despite the above findings, many important problems on the SWM phase are still unaddressed by the above works. First, even though some surface states are obtained by numerical ways, it is still unclear what is the fundamental topological reason that is responsible for their occurrence? Second, is there any bulk-surface correspondence in the SWM phase, and how the bulk topology determines the surface modes? Third, despite the obvious pairing difference between the internode BCS SWM and the intranode FFLO SWM, is there any underlying common characteristics that account for their similar surface states?

In this work, we answer the above three questions by deriving the effective TFT of the SWM phase. To do so, the key point would be to find out a correct external field that induces observable response behavior. Since the electromagnetic field is screened in the SC bulk, it is not considered as the suitable probe field. Hence, the gravitational or thermal field response theory is proposed in topological superconductors \cite{zhongwang,ryu}. However, if one is not interested in the bulk response but only focuses on the surface states, more feasible approaches are available without resorting to the coupling of the gravitational or thermal field. Different from Refs.\cite{zhongwang,ryu}, which investigate the bulk response of a general topological superconductor, we formulate in this work a simple and direct method to study the surface states of the SWM phase by constructing a $\mathrm{U}(1)$ TFT in the bulk. By coupling a $\mathrm{U}(1)$ gauge field to the SWM phase, one is able to arrive at the surface description that is independent of the external fields, since the surface states are only intrinsic manifestations of the bulk topology.  Following the above consideration, we obtain the following conclusions.  First, it is found that, after integrating out the matter fields, two 3+1D Chern-Simons-type actions emerge, describing the external $\mathrm{U}(1)$ gauge field. The first Chern-Simons term is due to the chiral anomaly, which is found to be robust against the onset of the superconductivity. It manifests itself by the Fermi arc of the SWM state. The second Chern-Simons term includes the contribution from a left-handed and a right-handed sector, and each sector further contains a particle-hole symmetric and a particle-hole antisymmetric term. These topological actions describe the $\mathrm{U}(1)$ gauge field coupled to the Bogoliubov quasi-particles, and are responsible for the SABSs of the SWM phase. Besides, the emergence of the such term is attributed to the second order renormalization  by the external gauge field, which has a similar counterpart in the quantum electrodynamics (QED), \textit{i.e.}, the parity anomaly. Therefore, we show that the surface states of the SWM phase are, in essence, the results of two quantum anomalies, i.e., the Fermi arc originates from the chiral anomaly and the SABS owes to the parity-like anomaly. Second, we have proved that, in the sense of TFT, the internode BCS state and the intranode FFLO state belong to the same universality class. Namely, they share the same effective field theory after the fermions are integrated out.

The remaining part of this work is organized as following. In Sec.\uppercase\expandafter{\romannumeral2}, we introduce the continue model of the SWM state for both the internode BCS pairing and the intranode FFLO pairing cases. In Sec.\uppercase\expandafter{\romannumeral3}, we briefly present our main method, where we also review the chiral anomaly of the WSM phase as an example. In Sec.\uppercase\expandafter{\romannumeral4}, the detailed analysis of the two quantum anomalies of the intranode FFLO SWM phase are discussed. In Sec.\uppercase\expandafter{\romannumeral5}, we calculate the surface states of the FFLO SWM with a finite boundary, which are exactly in agreement with the results from the TFT in the last section. In Sec.\uppercase\expandafter{\romannumeral6}, the internode BCS state is discussed. Finally, the conclusion and further discussion is presented in Sec.\uppercase\expandafter{\romannumeral7}.

\section{Model of the SWM}
To investigate the quantum anomalies, which are only dependent on the low-energy degrees of freedom near the Weyl points in momentum space while the higher energy window is irrelevant, we first start from a low-energy effective model of the simplest Weyl semimetals, where only two Weyl points (A,B) are present. The two Weyl points, with the opposite chirality, are separated in momentum space due to a TRS-breaking vector $\mathbf{Q}$.  A lattice model will be studied in section \uppercase\expandafter{\romannumeral5} to verify the results obtained by the continuum model.  Moreover, we consider a slight doping, such that the chemical potential $\mu$ can be higher (or lower) than the Weyl point. In this case, the low-energy physics is determined by two small spherical Fermi surfaces, which still enjoy the spin-momentum locking and the unique spin textures of the Weyl material. This state of matter, termed as the Weyl metals \cite{Bednik}, can be described by the following general Hamiltonian in the basis  $\Psi_{\mathbf{k}}=[c^A_{\mathbf{k}\uparrow},c^A_{\mathbf{k}\downarrow},c^B_{\mathbf{k}\uparrow},c^B_{\mathbf{k}\downarrow}]^T$,
\begin{equation}\label{eq1}
  H_0=\sum_{\mathbf{k}}\Psi^{\dagger}_{\mathbf{k}}[\tau^z\mathbf{k}\cdot\boldsymbol{\sigma}+\tau^0\mathbf{Q}\cdot\boldsymbol{\sigma}-\mu]\Psi_{\mathbf{k}}.
\end{equation}
In this work, we focus on the TRS-breaking SWM phase while neglecting the inversion-symmetry-breaking term in the above Hamiltonian. The vector $\mathbf{Q}$ breaks the TRS and separate the Weyl points in momentum space. Without losing any generality, we assume the Weyl points are split along $k_z$ by setting $\mathbf{Q}=(0,0,Q_z)$. For simplicity, we have assumed the isotropy of the Weyl points and set $v_F=1$. $\boldsymbol{\tau}$ and $\boldsymbol{\sigma}$ are the Pauli matrices to represent the chirality and spin degree of freedoms. Eq.\eqref{eq1} is a low-energy continuum model, and the sum of $\mathbf{k}$ should be within a momentum cutoff $\Lambda$ centered at each WP, i.e., $|\mathbf{k}\pm\mathbf{Q}|\leq\Lambda$. In the standard field theoretical treatment, $\Lambda$ can be regarded as infinity, which preserves all the Weyl physics in the low-energy window \cite{Zyuzin}.

Since the Fermi surfaces are now spherical pockets rather than discrete points due to the slight doping ($\mu\neq0$), the general state described by Eq.\eqref{eq1} is a metal. Different from the normal metals, the Weyl metal here exhibits the spin momentum locking and the band is nondegenerate. This property has motivated previous studies on superconductivity in the Weyl metal phase. Since our main aim in this work is the quantum anomalies and the intrinsic topological behavior of the SWM state, we do not discuss in detail the specific pairing symmetry and their energetics, but study only the topological response of two general possible pairing cases: the uniform intra-node pairing FFLO state and the uniform inter-node pairing BCS state.

Now in terms of the FFLO pairing state, the total Hamiltonian under our consideration is $H_1=H_{0}+H_{SC1}$. $H_{SC1}$ is the FFLO pairing term
\begin{equation}\label{eq2}
  H_{SC1}=\sum_{\mathbf{k}}\Delta c^{A\dagger}_{\mathbf{k}+\mathbf{Q}\uparrow}c^{A\dagger}_{-\mathbf{k}+\mathbf{Q}\downarrow}+\Delta c^{B\dagger}_{\mathbf{k}-\mathbf{Q}\uparrow}c^{B\dagger}_{-\mathbf{k}-\mathbf{Q}\downarrow}+H.C..
\end{equation}
Eq.\eqref{eq2} describes the intra-node pairing between electrons with momentum $\mathbf{k}+\mathbf{Q}$ and $-\mathbf{k}+\mathbf{Q}$ (for A node), which has a finite pairing momentum $2\mathbf{Q}$. In order to facilitate our following calculation, it is more convenient to write the superconductivity term in real space,
\begin{equation}\label{eq3}
  H_{SC1}=\sum_{\mathbf{r}}\Delta e^{-i 2\mathbf{Q}\cdot\mathbf{r}}c^{A\dagger}_{\mathbf{r}\uparrow}c^{A\dagger}_{\mathbf{r}\downarrow}+\Delta e^{i 2\mathbf{Q}\cdot\mathbf{r}} c^{B\dagger}_{\mathbf{r}\uparrow}c^{B\dagger}_{\mathbf{r}\downarrow}+H.C.,
\end{equation}
where we have treated the A, B chiralities as two different flavors that are independent of each other. Now one can introduce the Nambu space and the basis $\Phi_{\mathbf{r}}=[c^A_{\mathbf{r}\uparrow},c^A_{\mathbf{r}\downarrow},c^B_{\mathbf{r}\uparrow},c^B_{\mathbf{r}\downarrow},c^{B\dagger}_{\mathbf{r}\downarrow},c^{B\dagger}_{\mathbf{r}\uparrow},c^{A\dagger}_{\mathbf{r}\downarrow},c^{A\dagger}_{\mathbf{r}\uparrow}]^T$. In this basis, the total Hamiltonian reads,
\begin{equation}\label{eq4}
\begin{split}
  H_1=&\sum_{\mathbf{r}}\Phi^{\dagger}_{\mathbf{r}}[\tau^z\sigma^z(-i\partial_z)+Q_z\sigma^z+s^z\tau^z\sigma\cdot(-i\nabla_{\parallel})-\mu s^z \\
    &+s^x\tau^{+}\sigma^z\Delta e^{-2iQ_zz}+s^x\tau^{-}\sigma^z\Delta e^{2iQ_zz}]\Phi_{\mathbf{r}},
\end{split}
\end{equation}
where $\tau^{\pm}=\tau^x\pm i\tau^y$ and the Pauli matrix $\boldsymbol{s}$ is defined in the Nambu space. $\nabla_{\parallel}=(\partial_x,\partial_y)$ is the gradient operator in $x-y$ plane. In Eq.\eqref{eq4}, we have inserted $\mathbf{Q}=(0,0,Q_z)$ without the loss of generality.

To investigate the quantum anomalies in the above Hamiltonian Eq.\eqref{eq4}, it is natural to utilize the functional path-integral representation. Then, the imaginary-time action describing the FFLO superconducting Weyl metals can be written as,
\begin{equation}\label{eq5}
\begin{split}
  S_1&=\int d\tau d\mathbf{r} \Phi^{\dagger}_{r}[\partial_{\tau}+\tau^z\sigma^z(-i\partial_z)+s^z\tau^z\sigma\cdot(-i\nabla_{\parallel})\\
   &+\sigma^zQ_z-\mu s^z+s^x\tau^{+}\sigma^z\Delta e^{-2iQ_zz}+s^x\tau^{-}\sigma^z\Delta e^{2iQ_zz}]\Phi_{r}.
\end{split}
\end{equation}

Similarly, for the internode pairing BCS pairing state, we have  $H_{2}=H_{0}+H_{SC2}$, where the internode BCS pairing term $H_{SC2}$ reads
\begin{equation}\label{eq6}
  H_{SC2}=\sum_{\mathbf{k}}\Delta c^{A\dagger}_{\mathbf{k}\uparrow}c^{B\dagger}_{-\mathbf{k}\downarrow}+\Delta c^{B\dagger}_{\mathbf{k}\uparrow}c^{A\dagger}_{-\mathbf{k}\downarrow}+H.C..
\end{equation}
After making Fourier transformation back to the real space and using the functional path-integral formalism, we obtain the following imaginary-time action for the internode BCS superconducting Weyl metals,
\begin{equation}\label{eq7}
\begin{split}
  S_2&=\int d\tau d\mathbf{r} \Phi^{\dagger}_{r}[\partial_{\tau}+\tau^z\sigma^z(-i\partial_z)+s^z\tau^z\sigma\cdot(-i\nabla_{\parallel})\\
   &+\sigma^zQ_z-\mu s^z+s^x\sigma^z\Delta]\Phi_{r}.
\end{split}
\end{equation}

Eq.\eqref{eq5} and Eq.\eqref{eq7} are the models we are going to study. These models include the most important low-energy physics of the superconducting Weyl metals. Since Weyl semimetal enjoys the chiral anomaly which further leads to the semi-quantized quantum Hall effect and the chiral magnetic effect \cite{Zyuzin}, it is natural to ask the question whether the chiral anomaly would be preserved or not after the onset of superconductivity, and will there be any other quantum anomalies induced by the superconductivity? If such anomalies exist in superconducting Weyl metals, then what are their impacts on the surface boundaries? In what follows, we will give a detailed analysis on these problems.
\section{General method}
To answer the above questions, we are going to work out a topological field theory of the gauge field coupled to the SWM phase. In order to be more self-contained, we first briefly review the problem of the chiral anomaly in the WSM state, which can serve as general method in the following discussion on the superconducting Weyl metal phase.

For the simplest WSM state with two Weyl points, we start from the Eq.\eqref{eq1}. The general methods to study the topological response of WSM state can be concluded into three steps. First, resorting to the functional path-integral formalism, we arrive at the imaginary-time effective action,
\begin{equation}\label{eqq1}
  S_{wsm}=\int d\tau d\mathbf{r}\Psi^{\dagger}_{r}[\partial_{\tau}+\tau^z(-i\nabla)\cdot\boldsymbol{\sigma}+\tau^0\mathbf{Q}\cdot\boldsymbol{\sigma}-\mu]\Psi_{r},
\end{equation}

Second, we consider the coupling of a $\mathrm{U}(1)$ gauge field to Eq.\eqref{eqq1}.
\begin{equation}\label{eqq2}
  S=\int d\tau d\mathbf{r}\Psi^{\dagger}_{\mathbf{r},\tau}[\partial_{\tau}+ieA_0+\mathcal{H}(-i\mathbf{\nabla}+e\mathbf{A})]\Psi_{\mathbf{r},\tau},
\end{equation}
where $\mathcal{H}(-i\mathbf{\nabla}+e\mathbf{A})$ reads
\begin{equation}\label{eqq3}
  \mathcal{H}(-i\mathbf{\nabla}+e\mathbf{A})=\tau^z(-i\nabla+e\mathbf{A})\cdot\boldsymbol{\sigma}+\tau^0\mathbf{Q}\cdot\boldsymbol{\sigma}-\mu
\end{equation}
Third, we are ready to integrate out the matter fields. Before doing so, one can further simplify Eq.\eqref{eqq3} utilizing the symmetry of $S$. This trick has been used to obtain the gauge field theory associated with topological defects. For example, if there exists vortices in 2D, singular transformation can be performed to include the vortices degrees of freedom into the gauge field \cite{Seradjeh}. However, the symmetry of $S$ does not equal to the symmetry of the partition function $Z=\int Dc^{\dagger}Dce^{-S}$, \emph{i.e.}, the Jacobian associated to the transformation of the integral measure must be considered carefully, which may lead to physically measurable results. Therefore, the action can be formally written into
\begin{equation}\label{eqq6}
  S=\int d\tau d\mathbf{r}\Psi^{\dagger}_{\mathbf{r},\tau}[\partial_{\tau}+ieA_0+\widetilde{\mathcal{H}}(-i\mathbf{\nabla}+e\mathbf{A})]\Psi_{\mathbf{r},\tau}+\delta S,
\end{equation}
where $\widetilde{\mathcal{H}}(-i\mathbf{\nabla}+e \mathbf{A})$ denotes the transformed Hamiltonian and $\delta S$ represents the correction from the integral measure. Now the one fermion loop effective action for the gauge field can be obtained as
\begin{equation}\label{eqq7}
  S_{eff}[A_0,\mathbf{A}]=\mathrm{Tr}[\mathrm{log}(\partial_{\tau}+ieA_0+\widetilde{\mathcal{H}}(-i\mathbf{\nabla}+e\mathbf{A}))]+\delta S.
\end{equation}
A perturbative treatment of $A_{\mu}$ can further expand the above action, making possible the calculation of Feynman diagrams order by order. Eq.\eqref{eqq7} is the general effective action of the external gauge field, which include information of the topological response of the studied WSM state. It shows that the main results may come from two terms. The first term in Eq.\eqref{eqq7}  comes from the transformed Hamiltonian. For the WSM case, the $\widetilde{\mathcal{H}}$ after the chiral transformation describes a gapless Dirac fermion in the $\Gamma$ point. For gapless Dirac fermion, no topological nontrivial term would occur after calculating the trace \cite{Zyuzin}, therefore, the main contribution comes from the second term $\delta S$. Such term is due to the nonzero logarithm of the Jacobian of the functional integral measure under the chiral transformation, which breaks the chiral symmetry in the partition function even though it is preserved in the action. This phenomenon is termed as the chiral anomaly. The chiral anomaly of the WSM phase is well studied. By using regularization methods, such as the Fujikawa's method \cite{Fujikawa}, $\delta S$ is obtained as \cite{Zyuzin},
\begin{equation}\label{eqq8}
  \delta S=-\frac{e^2}{8\pi^2}\int d^3rdt\epsilon^{\mu\nu\rho\lambda}\partial_{\mu}\theta A_{\nu}\partial_{\rho}A_{\lambda},
\end{equation}
where $\theta$ is the axion angle determined by the separation of the Weyl points. This action clearly shows the emergence of the semi-quantized quantum Hall effect and the chiral magnetic effect.

The above analysis of the pure WSM state includes the main steps used in this work.
Different from the above pure WSM case, more issues need to be carefully considered after the onset of the superconductivity.  We can expect that the pairing gap will lead to a massive Dirac Hamiltonian after the chiral transformation. Hence, the first term in Eq.\eqref{eqq7} may also result in nontrivial topological field theory after the fermions are integrated out. This question on how the superconductivity would affect the topological response behavior of the WSM is not discussed in any literatures to the best of our knowledge, so we will show the derivation as detailed as possible. Moreover, the relationship between the emerging quantum anomalies and the surface states will be addressed via establishing the bulk-surface correspondence in the SWM.

\section{Quantum anomalies in FFLO state}
Now we first investigate the FFLO superconducting Weyl metal state described by the action Eq.\eqref{eq5}. The important symmetry in Eq.\eqref{eq5} is the chiral symmetry, therefore we first consider the local chiral transformation, $\Phi_{\mathbf{r}}\rightarrow e^{-is^0\tau^z\sigma^0\theta(\tau,\mathbf{r})/2}\Phi_{\mathbf{r}}$. We separate the total action $S$ into the Weyl metal term and the superconductivity term, \textit{i.e.}, $S=S_0+S_{SC1}$. As to $S_0$, performing the chiral transformation, $S_0$ is changed into $S^{\prime}_0$,
\begin{equation}\label{eq8}
\begin{split}
  S^{\prime}_0&=\int d\tau d\mathbf{r}\Phi^{\dagger}_{\mathbf{r}}[\partial_{\tau}+s^0\tau^z\sigma^z(-i\partial_z)+\sigma^z(Q_z-\frac{1}{2}\partial_z\theta)\\
  &-\mu s^z+s^z\tau^z\boldsymbol{\sigma}\cdot(-i\nabla_{\parallel})-\tau^z\frac{i}{2}\partial_{\tau}\theta
  -\frac{1}{2}s^z\boldsymbol{\sigma}\cdot\nabla_{\parallel}\theta]\Phi_{\mathbf{r}},
\end{split}
\end{equation}
The above equation suggests that the TRS-breaking term will be eliminated if $\theta(\tau,\mathbf{r})=2Q_zz$. This is an expected result since for pure Weyl metals, the Weyl nodes can be shifted, merging into a gapless Dirac point by the local transformation \cite{Zyuzin}. Now we consider the superconductivity terms. Under the local chiral transformation, we have
\begin{equation}\label{eq9}
\begin{split}
  S_{SC1}&\rightarrow S^{\prime}_{SC1}=\int d\tau d\mathbf{r}\Phi^{\dagger}_{\mathbf{r}}\\
  &\times\Delta[e^{-2iQ_zz+i\tau^z\theta} s^x\tau^{+}\sigma^z
  +e^{2iQ_zz+i\tau^z\theta} s^x\tau^{-}\sigma^z]\Phi_{\mathbf{r}}.
\end{split}
\end{equation}
After taking into account $e^{i\tau^z\theta}\tau^{\pm}=e^{\pm i\theta}\tau^{\pm}$ and $\theta=2Q_zz$, $S^{\prime}_{SC1}$ can be further simplified into
\begin{equation}\label{eq10}
   S^{\prime}_{SC1}=\int d\tau d\mathbf{r}\Phi^{\dagger}_{\mathbf{r}}\Delta s^x\tau^x\sigma^z\Phi_{\mathbf{r}}.
\end{equation}
Now the transformed total action $S^{\prime}_1=S^{\prime}_0+S^{\prime}_{SC1}$ reads in moment space,
\begin{equation}\label{eq11}
\begin{split}
  S^{\prime}&=\sum_{i\omega_n}\int\frac{d^3 k}{(2\pi)^3}\Phi^{\dagger}(i\omega_n,\mathbf{k})
  (-i\omega_n+\tau^z\sigma^zk_z-\mu s^z\\
  &+s^z\tau^z\sigma\cdot\mathbf{k}_{\parallel}+\Delta s^x\tau^x\sigma^z)\Phi(i\omega_n,\mathbf{k}).
\end{split}
\end{equation}
The above action is nontrivial due to its following properties. First, for $\mu\neq0$, since the chiral transformation shifts the two original Weyl points to $\Gamma$ point, the above action describes a well defined Dirac point at $\Gamma$ point, together with a superconducting gap below (or above) the Dirac point (see Fig.1(a)). Since we assume the chemical potential to be close enough to the Weyl node so that the spin-momentum locking is preserved in the small Fermi pocket, we can expect the pairing gap may show interesting topological properties. Second, for $\mu=0$, Eq.\eqref{eq11} depicts a fully gapped Dirac point at $\Gamma$ point (see Fig.1(b)). It is known that a gapless Dirac point is topologically trivial. Moreover, the gapped 2D Dirac fermion can be topologically nontrivial depending on the sign of the mass. Therefore, the 3D gapped Dirac fermion in Fig.1(b), if we view it as a layer of 2D case, may also enjoy nontrivial topological properties. Moreover, besides the transformed states in Fig.1, it is still unclear whether the chiral anomaly arising from the functional integral measure will be affected by the onset of superconductivity.
Hence, now we know that two aspects need to be addressed. One is the functional integral measure in the partition function, and the second one is the chiral-transformed state, Eq.\eqref{eq11}. In what follows, we will develop a general theory for both $\mu\neq0$ and $\mu=0$ case.  It is revealed that the nontrivial Jacobian $\delta S$ of the functional integral measure leads to robust Fermi arc in SWM, and the transformed action Eq.\eqref{eq11}, after calculating the radiative correction of the $U(1)$ gauge field, results in a Chern-Simons-like topological response in the one-loop effective action. This anomalous action, which has not been found in the SWM phases to the best of our knowledge, is the fundamental reason for the emerging in-gap SABSs.
\begin{figure}[tbp]
\includegraphics[width=3.4in]{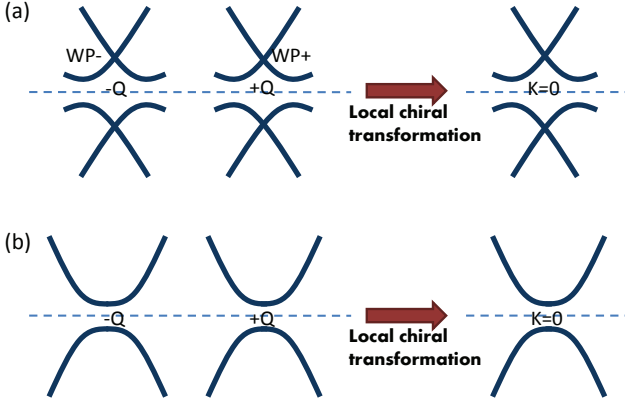}
\caption{(color online). The schematic diagram showing the change of the energy spectrum along $k_z$ of the intranode pairing FFLO SWM phase under the local chiral transformation. (a) is for the case chemical potential $\mu\neq0$. (b) shows the case $\mu=0$.}
\end{figure}
\subsection{The chiral anomaly}
Now let us investigate the first anomaly and extract the contribution $\delta S$. As the first step, a $U(1)$ gauge field $A_{\mu}$ (with $\mu=0,x,y,z$) needs to be introduced into Eq.\eqref{eq5}, which leads to
\begin{equation}\label{eq12}
\begin{split}
  S_1&=\int d\tau d\mathbf{r} \Phi^{\dagger}_{r}[\partial_{\tau}+ies^zA_0+\tau^z\sigma^z(-i\partial_z+es^zA_z)\\
  &+\sigma^zQ_z+s^z\tau^z\boldsymbol{\sigma}\cdot(-i\nabla_{\parallel}+es^z\mathbf{A}_{\parallel})
   -\mu s^z\\
   &+s^x\tau^{+}\sigma^z\Delta e^{-2iQ_zz}+s^x\tau^{-}\sigma^z\Delta e^{2iQ_zz}]\Phi_{r},
\end{split}
\end{equation}
where $A_{\parallel}=(A_x,A_y)$ are external electromagnetic field. When coupling gauge field to the superconductors where the Nambu space enlarges the original degrees of freedom, attention needs to be paid that the gauge field should not destroy the particle-hole symmetry, i.e., the gauge field terms that coupled to the hole states should be able to return back to the form of that coupled to the electron states. This leads to the $s^z$ in front of $A_{\mu}$ in Eq. \eqref{eq12}. Before we calculate the chiral anomaly, it is desirable to introduce first the Dirac matrices, $\gamma^{\mu}=is^0\tau^y\sigma^{\mu}$ with $\mu=1,2,3$, $\gamma^0=s^0\tau^x\sigma^0$ and $\gamma^4=-i\gamma^0$. A $\gamma^5$ matrix can be defined by $\gamma^5=-i\gamma^0\gamma^1\gamma^2\gamma^3=\tau^z$. With these definitions, $S_1$ can be written in a  more compact form,
\begin{equation}\label{eq13}
\begin{split}
  S_1&=\int d\tau d\mathbf{r} \overline{\Phi}_r\{i\gamma^4(\partial_0+is^3eA_0)+i\gamma^j(s^3\partial_j+ieA_j)\\
  &+i\gamma^3(\partial_3+is^3eA_3+i\gamma^5Q_3)-i\mu s^3\gamma^4\\\
  &+i\widetilde{\Delta}s^x(\frac{1-\gamma^5}{2})\gamma^5\gamma^4\gamma^3+i\widetilde{\Delta}^{\star}s^x(\frac{1+\gamma^5}{2})\gamma^5\gamma^4\gamma^3\}\Phi_r,
\end{split}
\end{equation}
where $j=1,2$, $\widetilde{\Delta}=-\Delta e^{-i2Q_zz}$ and $\overline{\Phi}=\Phi^{\dagger}\gamma^0$. The notations $(x,y,z)$ are set to be $(1,2,3)$ for brevity. Compared to the WSM case in Eq.\eqref{eqq2}, the action here is much more complicated due to the superconductivity. Writing the action into explicit matrix in the Nambu space, we can find the relationship between it and the action of the pure Weyl metal phase, which reads,
\begin{equation}\label{eq14}
  S=i\int d\tau d\mathbf{r}\overline{\Phi}_r\Theta\Phi_r,
\end{equation}
with $\Theta$ explicitly being written out as a 2 by 2 block matrix whose elements contain the Dirac matrices,
\begin{equation}\label{eq15}
  \Theta=\left(
           \begin{array}{cc}
             \Theta_{11} & \Theta_{12} \\
             \Theta_{21} & \Theta_{22} \\
           \end{array}
         \right),
\end{equation}
where
\begin{eqnarray}
                \Theta_{11} &=& \gamma^{\mu}(\partial_{\mu}+ieA_{\mu}+i\gamma^5Q_{\mu})-\mu\gamma^4, \\
                \Theta_{12} &=& (\widetilde{\Delta}\frac{\gamma^5-1}{2}+\widetilde{\Delta}^{\star}\frac{\gamma^5+1}{2})\gamma^4\gamma^3, \\
                \Theta_{21} &=& (\widetilde{\Delta}\frac{\gamma^5-1}{2}+\widetilde{\Delta}^{\star}\frac{\gamma^5+1}{2})\gamma^4\gamma^3,\\
                \Theta_{22} &=& \widetilde{\gamma}^{\mu}(\partial_{\mu}-ieA_{\mu}+i\gamma^5Q_{\mu})+\mu\gamma^4,
              \end{eqnarray}
where we have redefined $\widetilde{\gamma}^{\mu}=(-\gamma^1,-\gamma^2,\gamma^3,\gamma^4)$. As expected, $\Theta_{11}$ exactly equals to the action in non-superconducting Weyl metals, this term can lead to the chiral anomaly when the function integral measure is studied carefully. Besides this term, $\Theta_{22}$ is the hole-version action, which is the particle-hole transformation of $\Theta_{11}$. Therefore, we expect this term also results in the chiral anomaly. The off-diagonal terms describes the pairing contributions, whose effect will be analyzed in what follows.

To calculate the Jacobian of the functional integral measure in the partition function, $Z=\int D\overline{\Phi}D\Phi e^{S}$, we use an abstract notation $\hat{U}$ to denote the chiral transformation $\Phi_{\mathbf{r}}\rightarrow e^{-is^0\tau^z\sigma^0\theta(\tau,\mathbf{r})/2}\Phi_{\mathbf{r}}$. Then, the Jacobian $J$ under the chiral transformation $\hat{U}$ can be calculated as $J=Det(U^{-2})$. In order to calculate the determinant, an appropriate basis needs to be clarified first so that one can obtain the specific form of $\hat{U}$ in this basis. The most natural basis would be the eigenvector of the operator $\Theta$, however, one can easily find that $\Theta$ is not hermitian, which does not have eigenvectors corresponding to real eigenvalues. Therefore, the first step would be searching for hermitian operators based on $\Theta$. The simplest ones are $\Xi_1=\Theta^{\dagger}\Theta$ and $\Xi_2=\Theta\Theta^{\dagger}$. Using these hermitian operators, now we can have well-defined basis so that the Jacobian can be finally obtained. Before showing the detailed calculation, we can write down the following eigenvalue equations.
\begin{eqnarray}
  \Xi_1\varphi_n &=& \epsilon_n\varphi_n, \\
  \Xi_2\varphi^{\prime}_n &=& \epsilon_n\varphi^{\prime}_n.
\end{eqnarray}
$\varphi_n$ and $\varphi^{\prime}_n$ are the eigenvectors of $\Xi_1$ and $\Xi_2$, respectively. $\varphi_n$ ($\overline{\varphi}_n$) and $\varphi^{\prime}_n$ ($\overline{\varphi^{\prime}}_n$) lie in the linear spaces that are isomorphic to the linear space constructed by $\Phi_{r}$ ($\overline{\Phi}_r$). Therefore, we can expand the Grassmann $\Phi_{r}$ (and $\overline{\Phi}_{r}$) using $\varphi_n$ and $\varphi^{\prime}_n$, \textit{i.e.},
\begin{eqnarray}
  \Phi_{r} &=& \sum_n a_n\varphi_n(r)= \sum_n a^{\prime}_n\varphi^{\prime}_n(r) \\
  \overline{\Phi}_{r} &=& \sum_n a^{\star}_n\overline{\varphi}_n(r)= \sum_n a^{\prime\star}_n\overline{\varphi^{\prime}}_n(r).
\end{eqnarray}
Now we consider the chiral transformation $\hat{U}$, under which we have, $\hat{U}\Phi_r=\Phi^{\prime}_{r}$. Since $\hat{U}$ acts directly on $\Phi_r$, and $\Phi_r$ can be expanded by $\varphi_n$ or $\varphi^{\prime}_n$, we can now write $\hat{U}$ in the basis of $\varphi_n$ or $\varphi^{\prime}_n$, which reads,
\begin{eqnarray}
  U_{mn} &=& e^{-\frac{i}{2}\int d^4x\varphi^{\star}_n(x)s^0\gamma^5\theta(x)\varphi_m(x)} \\
  U^{\prime}_{mn} &=& e^{-\frac{i}{2}\int d^4x\varphi^{\prime\star}_n(x)s^0\gamma^5\theta(x)\varphi^{\prime}_m(x)},
\end{eqnarray}
respectively. Then, the Jacobian transformation is obtained in these basis as,
\begin{equation}\label{eq16}
  J=e^{\frac{i}{2}(P_1+P_2)},
\end{equation}
where
\begin{eqnarray}
  P_1 &=& \int d^4x P_1(x)=\sum_n\int d^4x\varphi^{\star}_n(x)s^0\gamma^5\theta(x)\varphi_n(x), \\
  P_2 &=&  \int d^4x P_2(x)=\sum_n\int d^4x\varphi^{\prime\star}_n(x)s^0\gamma^5\theta(x)\varphi^{\prime}_n(x).
\end{eqnarray}
To evaluate $P_1$ and $P_2$ explicitly, the standard method of heat kernel regularization is always used \cite{Fujikawa,Zyuzin}, \textit{i.e.},
\begin{eqnarray}
  P_1(x) &=& \lim_{M\rightarrow\infty}\sum_n\varphi^{\star}_n(x)s^0\gamma^5\theta(x)e^{\frac{-\Theta^{\dagger}\Theta}{M^2}}\varphi_n(x), \\
  P_2(x) &=& \lim_{M\rightarrow\infty}\sum_n\varphi^{\prime\star}_n(x)s^0\gamma^5\theta(x)e^{\frac{-\Theta\Theta^{\dagger}}{M^2}}\varphi^{\prime}_n(x).
\end{eqnarray}
Making expansion using the plane waves, the above equations are transformed to a more feasible form
\begin{eqnarray}
  P_1(x) &=& (P^{11}_1(x)+P^{22}_1(x))/2,\\
  P_2(x) &=& (P^{11}_2(x)+P^{22}_2(x))/2,
\end{eqnarray}
where the factor $1/2$ is due to the fact that the Nambu space doubles the physical degrees of freedom, and the redundancy should be removed. In the above equation, we have
\begin{eqnarray*}
  P^{11}_1(x) &=& \theta\lim_{M\rightarrow\infty}\int \frac{d^4k}{(2\pi)^4}\mathrm{tr}_{\gamma}[\gamma^5e^{-ikx}e^{-\Xi_1/M^2}e^{ikx}]^{11}, \\
  P^{22}_1(x) &=& \theta\lim_{M\rightarrow\infty}\int \frac{d^4k}{(2\pi)^4}\mathrm{tr}_{\gamma}[\gamma^5e^{-ikx}e^{-\Xi_1/M^2}e^{ikx}]^{22}, \\
  P^{11}_2(x) &=& \theta\lim_{M\rightarrow\infty}\int \frac{d^4k}{(2\pi)^4}\mathrm{tr}_{\gamma}[\gamma^5e^{-ikx}e^{-\Xi_2/M^2}e^{ikx}]^{11}, \\
  P^{22}_2(x) &=& \theta\lim_{M\rightarrow\infty}\int \frac{d^4k}{(2\pi)^4}\mathrm{tr}_{\gamma}[\gamma^5e^{-ikx}e^{-\Xi_2/M^2}e^{ikx}]^{22},
\end{eqnarray*}
where the notation $^{11}$ ($^{22}$) denotes the $[1,1]$ ($[2,2]$) elements in the Nambu space and $\mathrm{tr}_{\gamma}$ represents the trace of the Dirac matrices. To calculate the above expression, we can expand the exponential $e^{-\Xi_1/M^2}$. Since in the heat kernel regularization formalism, we will finally rescale the momentum $k_{\mu}\rightarrow Mk_{\mu}$ \cite{Fujikawa}, therefore, only terms that contains $M^{-4}$ will survive and give a constant value under the limit $M\rightarrow\infty$. This means that we only need to consider the expansion of the exponential up to the second order. Moreover, for the zero and first order expansion, the trace over the Dirac matrices automatically gives zero, which further simplify the evaluation to only the second order expansion, which leads to
\begin{eqnarray}
  P^{11}_1(x) &=& \frac{\theta}{2}\int \frac{d^4k}{(2\pi)^4}\mathrm{tr}_{\gamma}[\gamma^5e^{-ikx}\Pi^{11}_1e^{ikx}] \\
  P^{22}_1(x) &=& \frac{\theta}{2}\int \frac{d^4k}{(2\pi)^4}\mathrm{tr}_{\gamma}[\gamma^5e^{-ikx}\Pi^{22}_1e^{ikx}]
\end{eqnarray}
where we have rescaled the momentum. $P^{11}_2(x)$ and $ P^{22}_2(x)$ can be similarly obtained, which are not shown for brevity. In the second order expansion above, we encountered the operators $\Pi^{11}_1=(\Theta^{\dagger}\Theta\Theta^{\dagger}\Theta)^{11}$ and $\Pi^{22}_1=(\Theta^{\dagger}\Theta\Theta^{\dagger}\Theta)^{22}$. After a straightforward calculation, $\Pi^{11}_1$ can be obtained as ($\Pi^{22}_1$ can be calculated in the same way, which is not shown explicitly),
\begin{equation}\label{eq18}
\begin{split}
  \Pi^{11}_1&=[(D_1^{\prime}+\mu\gamma^4)(D_1-\mu\gamma^4)+|\widetilde{\Delta}|^2]^2\\
  +&\{(D^{\prime}_1+\mu\gamma^4)(\widetilde{\Delta}\frac{\gamma^5-1}{2}+\widetilde{\Delta}^{\star}\frac{\gamma^5+1}{2})\gamma^4\gamma^3\\
  &-(\widetilde{\Delta}^{\star}\frac{\gamma^5-1}{2}+\widetilde{\Delta}\frac{\gamma^5+1}{2})\gamma^4\gamma^3(D_2+\mu\gamma^4)\}\\
  &\times\{(\widetilde{\Delta}^{\star}\frac{\gamma^5-1}{2}+\widetilde{\Delta}\frac{\gamma^5+1}{2})\gamma^3\gamma^4(D_1-\mu\gamma^4)\\
  &-(D^{\prime}_2-\mu\gamma^4)(\widetilde{\Delta}\frac{\gamma^5-1}{2}+\widetilde{\Delta}^{\star}\frac{\gamma^5+1}{2})\gamma^3\gamma^4\},
\end{split}
\end{equation}
where we have introduced notations $D_1=\gamma^{\mu}(\partial_{\mu}+ieA_{\mu}+i\gamma^5Q_{\mu})$, $D^{\prime}_1=\gamma^{\mu}(\partial_{\mu}+ieA_{\mu}-i\gamma^5Q_{\mu})$, $D_2=\widetilde{\gamma}^{\mu}(\partial_{\mu}-ieA_{\mu}+i\gamma^5Q_{\mu})$ and $D^{\prime}_2=\widetilde{\gamma}^{\mu}(\partial_{\mu}-ieA_{\mu}-i\gamma^5Q_{\mu})$ for brevity. Now we need to search for the nontrivial terms after inserting Eq.\eqref{eq18} into $P^{11}_1(x)$. Then after using Eq.\eqref{eq16}, we can obtain the anomalous topological response for the superconducting Weyl metal phase. Most terms vanish due to the trace over the Dirac matrices. Recalling the relation $tr_{\gamma}(\gamma^5\gamma^{\mu}\gamma^{\nu}\gamma^{\rho}\gamma^{\sigma})=-4\epsilon^{\mu\nu\rho\sigma}$ , it is known that the only nontrivial terms from Eq.\eqref{eq18} are those that enjoy four different Dirac matrices (except for $\gamma^5$). The superconducting terms that contains $\widetilde{\Delta}$ or $\widetilde{\Delta}^{\star}$ do not satisfy this condition, therefore they leads to vanishing contribution to the topological behavior. After checking every term, it is found that only the following term may survive and give a nontrivial contribution, so that Eq.\eqref{eq18} can be simplified as,
\begin{equation}\label{eq19}
\Pi^{11}_1=(D^{\prime}_1+\mu\gamma^4)(D_1-\mu\gamma^4)(D^{\prime}_1+\mu\gamma^4)(D_1-\mu\gamma^4).
\end{equation}
The expansion of the above equation further leads to sixteen terms, where the terms that have more than two $\gamma^4$ leads to zero after the trace together with $\gamma^5$. There are four terms with only one $\gamma^4$, whose absolute values are given by
\begin{equation}\label{eq20}
 P^{11}_1(x)=\frac{\theta\mu}{2}\int\frac{d^4k}{(2\pi)^4}\mathrm{tr}_{\gamma}[\gamma^5\gamma^4e^{-ikx}D_1D^{\prime}_1D_1e^{ikx}].
\end{equation}
However, due to the different signs in front of $\mu\gamma^4$ in Eq.\eqref{eq19}, the four terms exactly cancel with each other.
Hence, we know that the only possible contribution would come from the term with zero $\gamma^4$ matrix in Eq.\eqref{eq19}, which, after inserting into Eq.(39), reads
\begin{equation}\label{eq22}
  P^{11}_1(x)=\frac{\theta}{2}\int\frac{d^4k}{(2\pi)^4}\mathrm{tr}_{\gamma}[\gamma^5e^{-ikx}D^{\prime}_1D_1D^{\prime}_1D_1e^{ikx}].
\end{equation}
In the above equation, the trace of the Dirac matrices leads to the Levi-Civita tensor $\epsilon^{\mu\nu\rho\sigma}$. Further using the anti-symmetricity of $\epsilon^{\mu\nu\rho\sigma}$, and taking into account the relation $[\partial_{\mu}+ieA_{\mu},\partial_{\nu}+ieA_{\nu}]=ieF_{\mu\nu}$, the $D^{\prime}_1$ and $D_1$ operators can be reorganized into the electromagnetic tensor $F_{\mu\nu}$ of external gauge field. Then, we arrive at
\begin{equation}\label{eq23}
  P^{11}_1(x)=-\frac{e^2}{32\pi^2}\theta(x)\epsilon^{\mu\nu\rho\sigma}F_{\mu\nu}F_{\rho\sigma}.
\end{equation}
In terms of $P^{22}_1(x)$, following the same procedure before, we have
\begin{equation}\label{eq24}
  P^{22}_1(x)=\frac{\theta}{2}\int\frac{d^4k}{(2\pi)^4}\mathrm{tr}_{\gamma}[\gamma^5e^{-ikx}D^{\prime}_2D_2D^{\prime}_2D_2e^{ikx}],
\end{equation}
where $D_2$ and $D^{\prime}_2$ both contain the Dirac matrix $\widetilde{\gamma}^{\mu}$ instead of $\gamma^{\mu}$, as in $D_1$ and $D^{\prime}_1$. However, the trace over the Dirac matrix still satisfy $tr_{\gamma}(\gamma^5\widetilde{\gamma^{\mu}}\widetilde{\gamma^{\nu}}\widetilde{\gamma^{\rho}}\widetilde{\gamma^{\sigma}})=-4\epsilon^{\mu\nu\rho\sigma}$, since the two minus signs cancel with each other in the trace. Therefore, we have
 \begin{equation}\label{eq25}
  P^{22}_1(x)= P^{11}_1(x)=-\frac{e^2}{32\pi^2}\theta(x)\epsilon^{\mu\nu\rho\sigma}F_{\mu\nu}F_{\rho\sigma}.
\end{equation}
$P^{22}_1(x)$ and $P^{11}_1(x)$ describe the contribution of the hole and the particle degrees of freedom, respectively. Their equivalence between them is a result of the particle-hole symmetry in the SWM state.

Now, we can evaluate $P^{11}_2(x)$ and $P^{22}_2(x)$ exactly in the same way. A straightforward calculation reveals that $P^{11}_2(x)=P^{22}_2(x)=P^{22}_1(x)=P^{11}_1(x)$. After inserting every term into Eq.\eqref{eq16}, one will find that the Jacobian $J$ associated with the chiral transformation results in an additional action $\delta S$,
\begin{equation}\label{eq26}
  \delta S=-\frac{e^2}{32\pi^2}\int d^4 x\theta(x)\epsilon^{\mu\nu\alpha\beta}F_{\mu\nu}F_{\alpha\beta}.
\end{equation}
This action arises from the chiral anomaly, and it has the same form with that of the WSM phase \cite{Zyuzin}. Therefore, the first conclusion can be drawn up to now. Even though we consider a completely different state from WSM phase, \textit{i.e.}, the FFLO superconducting state of doped WSM phase, the field theoretical analysis shows that the chiral anomaly is still preserved. The superconducting gap does not affect the original topological response. 
As will be discussed in contents below, the action $\delta S$ shows the existence of a surface state in the SWM with a boundary. For the SWM phase, the surface state can be understood as the Fermi arc after projection to zero energy \cite{Wan}, therefore, what we show here indicates that the Fermi arc of the original WSM phase remains robust after the superconductivity sets in. This is reasonable, since superconductivity, which is the pairing state of the bulk electron, may hardly affect the state in the surface (at least is true for the uniform s-wave case studied here).

Besides, the above theory also reveals an interesting observation. As shown in the previous detailed calculation, even though we set an arbitrary parameter for the chemical potential (it is assumed to be small), the final result of the topological response does not rely on it at all. For an ideal case, where $\mu=0$ (in this case, the intrinsic superconductivity may be difficult to set in, but it can be induced by proximity effect), it is found that the FFLO pairing gaps out the Weyl nodes, leading to a full gap in the whole Brillouin zone. However, the action Eq.\eqref{eq26} can still be obtained in the same way. This suggests a remarkable conclusion that the Fermi arc of the Weyl metal phase can be preserved even though all the Weyl points are destroyed. Such prediction is unexpected, since the previous works tend to believe that the Fermi arc is the result of the Weyl nodes in the bulk. Our analysis indicates that the Fermi arc as well as the quantum anomalous Hall effect is not a direct consequence of the Weyl points. The Fermi arc will be preserved as long as the axion field theory Eq.\eqref{eq26} is not destroyed, or in other words, as long as the Berry curvature remains to be nontrivial around the Weyl valleys. To further confirm the above conclusions, in Sec.\uppercase\expandafter{\romannumeral6}, we will demonstrate the bulk-surface correspondence and derive the action of the surface state. Also, the robustness of the Fermi arc in the SWM will be revisited by studying a tight-binding model.

\subsection{The parity-like anomaly}
In this subsection, we are going to show that, besides the chiral anomaly discussed in the above section, there is an another quantum anomaly arising after the development of the superconductivity. This anomaly is a direct consequence of the superconducting gap, and can be revealed in the first term in Eq.\eqref{eqq6}. For the FFLO SMW phase, the first term of Eq.\eqref{eqq6} is explicitly written out as Eq.\eqref{eq11}, which describes the effective action of the FFLO SMW state after the local chiral transformation (see Fig.1).

Now we are going to show that the action Eq.\eqref{eq11} is also nontrivial after considering the one-loop radiative correction of the fermions. Before showing the detailed derivation, let us now consider a more realistic case where $\Delta\ll\mu$, since in realistic materials the pairing gap is usually much smaller than the chemical potential \cite{ruisec}. As the first step, we perform the canonical transformation $s^{\pm}\rightarrow\sigma^z s^{\pm}$, $\sigma^{\pm}\rightarrow s^z\sigma^{\pm}$, so that the action is simplified as
\begin{equation}\label{eq27}
\begin{split}
  S^{\prime}&\rightarrow\int d\tau d\mathbf{r}\Phi^{\prime\dagger}_{r}\{\partial_{\tau}+\tau^z\sigma^z(-i\partial_z)+\tau^z\boldsymbol{\sigma}\cdot(-i\nabla_{\parallel}),\\
  &+\Delta s^x\tau^x-\mu s^z\}\Phi^{\prime}_{r},
\end{split}
\end{equation}
where $\Phi^{\prime\dagger}_r$, $\Phi^{\prime}_r$ are the transformed eight dimensional spinors. Then, it is useful to introduce a set of new Dirac matrices $\eta^{\mu}$. Different from the Dirac matrices $\gamma^{\mu}$, now it is more convenient to set $\eta^i=\sigma^i$ ($i=1,2,3$), $\eta^0=\sigma^0$ and $\eta^4=i\eta^0$, with $\sigma^{\mu}$ being the Pauli matrices denoting the spin degrees of freedom. Then, writing explicitly in the chirality space, we arrive at
\begin{equation}\label{eq28}
  S^{\prime}=\int d\tau d\mathbf{r}\overline{\Phi}^{\prime}_r
  \left(
  \begin{array}{cc}
  -i\eta^{\mu}\partial_{\mu}+i\mu s^3\eta^4 & -i\Delta s^x\eta^4 \\
  -i\Delta s^x\eta^4 & -i\widetilde{\eta}^{\mu}\partial_{\mu}+i\mu s^3\eta^4 \\
  \end{array}
  \right)
  \Phi^{\prime}_r,
\end{equation}
where $\overline{\Phi}^{\prime}_r=\Phi^{\prime\dagger}_r\eta^0$, and $\widetilde{\eta}^{\mu}=(-\eta^1,-\eta^2,-\eta^3,\eta^4)$. Now we can further make Bogoliubov transformation to diagonalize the Nambu space. In momentum space, we obtain a simple expression for the SWM state after the local chiral transformation,
\begin{equation}\label{eq29}
  S^{\prime}=\int_{\Lambda}\frac{d^4k}{(2\pi)^4}\overline{\Phi}^{\prime\prime}_k \Omega(k)\Phi^{\prime\prime}_k,
\end{equation}
with $\Omega(k)$ being the Dirac kernel,
\begin{equation}\label{eq30}
\Omega(k)=\left(
\begin{array}{cc}
s^0\slashed{k}+s^3m & 0 \\
0 & s^0\widetilde{\slashed{k}}+s^3m \\
\end{array}
\right),
\end{equation}
where $m=\mu+\Delta^2/2\mu$. $\overline{\Phi}^{\prime\prime}_k$ and $\Phi^{\prime\prime}_k$ are the Grassmann fields describing the Bogoliubov quasi-particles.  The Feynman notation $\slashed{k}=\eta^{\mu}k_{\mu}$ and $\widetilde{\slashed{k}}=\widetilde{\eta}^{\mu}k_{\mu}$ are used. In deriving Eq.\eqref{eq30}, we have neglected the higher order term $O(\Delta^2)$, and obtained a linear Dirac fermion model in the momentum space. Moreover, a momentum cutoff $\Lambda=(\mu-\Delta)$ is required in the integration of $k$ due to the SC gap. After this treatment, we can discard the irrelevant degrees of freedom that are far from WP, while keep all the important physics, including the spin-orbit coupled Dirac cone as well as the SC gap.

Now we are ready to couple a $U(1)$ gauge field to the SWM state and then evaluate its one fermion loop effective action, \textit{i.e.},
\begin{equation}\label{eq31}
  S^{\prime}_{eff}[A_{\mu}]=\mathrm{Tr}\log(\Omega(k)+\Omega(A_{\mu})),
\end{equation}
with $\Omega(A_{\mu})$ being
\begin{equation}\label{eq32}
  \Omega(A_{\mu})=\left(
\begin{array}{cc}
s^0\slashed{A}& 0 \\
0 & s^0\widetilde{\slashed{A}} \\
\end{array}
\right),
\end{equation}
where $\slashed{A}=\eta^{\mu}A_{\mu}$ and $\widetilde{\slashed{A}}=\widetilde{\eta}^{\mu}A_{\mu}$. Here, $A_{\mu}$ is not the electromagnetic field, but is a gauge field formally introduced to couple to the Bogoliubov Grassmann fields. This treatment will facilitate us to obtain all possible surface states through the bulk-surface correspondence (see below). Now we treat $A_{\mu}$ perturbatively, expanding the logarithm in Eq.\eqref{eq31} in terms of $\Omega(A_{\mu})$. For the zero and first order expansion, no topologically nontrivial term can be obtained. In the following, we will show the detailed calculation in the dominant leading order that gives us the nontrivial topological properties. The second order process is shown by the Feynman diagram in Fig.2. The renormalized effective action in this order reads,
\begin{figure}[tbp]
\includegraphics[width=3.4in]{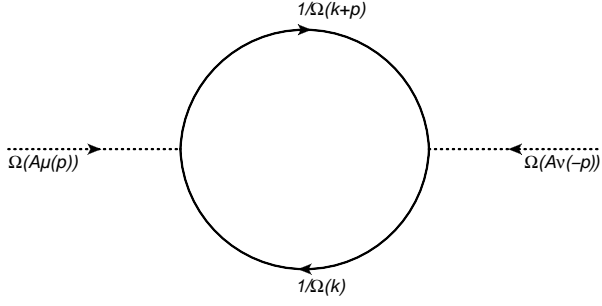}
\caption{(color online). The Feynman diagram in the second order perturbative calculation. The straight lines denote the propagator of the Grassmann field $\Phi^{\prime\prime}_{k}$ and the dashed lines represent the propagator of the gauge field.}
\end{figure}
\begin{equation}\label{eq33}
\begin{split}
  S^{\prime}_{eff}&=\frac{1}{2}\mathrm{Tr}[\Omega^{-1}\Omega(A_{\mu})\Omega^{-1}\Omega(A_{\mu})]
\end{split}
\end{equation}
After some straightforward algebra, the effective action can be calculated as
\begin{equation}\label{eq34}
  S^{\prime}_{eff}=\frac{1}{2}\int\frac{d^4pd^4k}{(2\pi)^8}\mathrm{tr}[\hat{f}_{k,p}(s^3)\mathbf{M}(k,p,A_{\mu})],
\end{equation}
where $\mathbf{M}(k,p,A_{\mu})$ is the matrix
\begin{equation}\label{eqa35}
  \mathbf{M}(k,p,A_{\mu})=\left(
                             \begin{array}{cc}
                               M_{11}(s^3) & 0 \\
                               0 & M_{22}(s^3) \\
                             \end{array}
                           \right),
\end{equation}
with \begin{eqnarray*}
       M_{11}(s^3) &=& (\widetilde{\slashed{k}}+s^3m)\slashed{A}_p((\widetilde{\slashed{k}}+\widetilde{\slashed{p}})+s^3m)\slashed{A}_{-p} \\
       M_{22}(s^3) &=& (\slashed{k}+s^3m)\widetilde{\slashed{A}}_p((\slashed{k}+\slashed{p})+s^3m)\widetilde{\slashed{A}}_{-p}.
     \end{eqnarray*}
In Eq.\eqref{eq34}, $\hat{f}(k,p)$ reads
\begin{equation}\label{eq35}
\hat{f_{k,p}}(s^3)=\frac{1}{(-k^2+m^2+i2s^3k_4m)(-k^2_p+m^2+i2s^3k_{p4}m)},
\end{equation}
where $k_p=k+p$. From Eq.(56), we can decompose the total action into four sectors, $S^{\prime}_{eff}=\sum_{i,j=\pm}S^{\prime}_{eff}[i,j]$, with $S^{\prime}_{eff}[\pm,\pm]$ being defined as following,
\begin{eqnarray}
  S^{\prime}_{eff}[+,+] &=& \int\frac{d^4pd^4k}{2(2\pi)^8}f_{k,p}(+)tr_{\eta}M_{11}(+), \\
  S^{\prime}_{eff}[+,-] &=& \int\frac{d^4pd^4k}{2(2\pi)^8}f_{k,p}(-)tr_{\eta}M_{11}(-), \\
  S^{\prime}_{eff}[-,+] &=& \int\frac{d^4pd^4k}{2(2\pi)^8}f_{k,p}(+)tr_{\eta}M_{22}(+), \\
  S^{\prime}_{eff}[-,-] &=& \int\frac{d^4pd^4k}{2(2\pi)^8}f_{k,p}(-)tr_{\eta}M_{22}(-),
\end{eqnarray}
where $\mathrm{tr}_{\eta}$ denotes the trace of the Dirac matrices $\eta^{\mu}$. $S^{\prime}_{eff}[+,+]$ comes from the component where $\tau^z=1$ and $s^z=1$, therefore, it is the contribution from the left-handed particle-hole symmetric degrees of freedom. Similarly, $S^{\prime}_{eff}[+,-]$, $S^{\prime}_{eff}[-,+]$, $S^{\prime}_{eff}[-,-]$ are the contribution from the left-handed particle-hole antisymmetric, the right-handed particle-hole symmetric and the right-handed particle-hole antisymmetric Bogoliubov quasi-particles, respectively.

Now, as an example, we search for the topological-nontrivial action of the gauge-field (Chern-Simons-like) in $S^{\prime}_{eff}[+,+]$. First, in the term $M_{11}(+)$, only the combination with three different Dirac matrices (except for $\eta^4$) does not vanish and gives us a Levi-Civita tensor after the trace. Second, for any Chern-Simons-like term to occur in the gauge field action, the gradient of the vector potential should be present, which means that only terms with the operator $\widetilde{\slashed{p}}$ can give us the topological-nontrivial response of the SWM state. With the above consideration, it is straightforward to arrive at the first topological anomalous action, which comes from the combination of
$m\slashed{A}_p\widetilde{\slashed{p}}\slashed{A}_{-p}$ in $M_{11}(+)$,
\begin{equation}\label{eq36}
  S^{\prime}_{eff}[+,+]=-m\int\frac{d^4p}{(2\pi)^4}\epsilon^{ijk}C(p)A_{i}(p)p_{j}A_{k}(-p),
\end{equation}
where $i,j,k=1,2,3$. The function $C(p)$ comes from the integral
\begin{equation}\label{eq37}
  C(p)=i\int_{\Lambda}\frac{d^4k}{(2\pi)^4}\hat{f_{k,p}}(+),
\end{equation}
which is free from the ultraviolet divergence due to the cutoff. In deriving Eq.\eqref{eq36}, the trace over the Dirac matrices $\eta^{\mu}$ needs careful consideration since the case is completely different from the usual four by four Dirac matrices. The important trace one encounters is $tr(\eta^{\mu}\eta^{\nu}\eta^{\rho})$, with $\mu,\nu,\rho=1,2,3,4$. It is easy to note that the trace vanish when any of the Dirac matrix has the notation 4 due to $tr(\eta^4\eta^{\mu}\eta^{\nu})=0$ (except for the case $\mu=\nu=\rho=4$, which gives a topological trivial term), therefore the $\eta^4$ term should be removed in order to have nonzero result. This leads to $tr(\eta^{i}\eta^{j}\eta^{k})=2i\epsilon^{ijk}$, with $i,j,k=1,2,3$.

Finally, in the real space, we obtain from Eq.\eqref{eq36} that
\begin{equation}\label{eq38}
  S^{\prime}_{eff}[+,+]=-m\int d^4xd^4x^{\prime}\epsilon^{ijk}C(x-x^{\prime})A_{i}(x)\partial_jA_k(x^{\prime}),
\end{equation}
where we have performed the Wick-rotation back to the real time space.  Now we can assume the gauge field has a long wave length compared to the SWM sample. In this long wave length limit $p\rightarrow0$, the integration in Eq.\eqref{eq37} is simplified to a constant $C_1=C(p\rightarrow0)$ \cite{Peskill}. Then, Eq.\eqref{eq38} becomes,
\begin{equation}\label{eqn38}
  S^{\prime}_{eff}[+,+]=-mC_1\int d^4x\epsilon^{ijk}A_{i}(x)\partial_jA_k(x),
\end{equation}
The above equation is clearly a Chern-Simon action of the external gauge field. To reveal the properties of the action Eq.\eqref{eq38}, we can take a variation of the external gauge field $A_{\mu}$, which gives us the topological response of the SWM state as following
\begin{equation}\label{eq39}
  j^{i}(x)=-mC_1\epsilon^{ijk}\partial_jA_k(x)=-mCB^i(x),
\end{equation}
which suggests that a current $j^i$ will be induced along the applied external field $B^i$ in the bulk. This looks similar to the chiral magnetic effect of the WSM material. However, one can find that this bulk response will be canceled exactly and leads to zero current, after the $M_{11}(-)$ term is taken into account. This can be easily shown by inserting $s^3=-1$ into $M_{11}(s^3)$, then we have
\begin{equation}\label{eq40}
  S^{\prime}_{eff1}[+,-]=mC_1\int d^4x\epsilon^{ijk}A_{i}(x)\partial_jA_k(x).
\end{equation}
This term induces a chiral magnetic effect that exactly cancels with Eq.\eqref{eq39}, therefore no response can be observed in the SWM bulk. Similarly, we can obtain $S^{\prime}_{eff}[-,+]$ and $S^{\prime}_{eff}[-,-]$ as,
\begin{eqnarray}
  S^{\prime}_{eff}[-,+] &=& mC_1\int d^4x\epsilon^{ijk}A_{i}(x)\partial_jA_k(x), \\
  S^{\prime}_{eff}[-,-] &=& -mC_1\int d^4x\epsilon^{ijk}A_{i}(x)\partial_jA_k(x).
\end{eqnarray}

Then, let us check the second possibility, i.e., the combination of terms $\widetilde{\slashed{k}}\slashed{A}_p\widetilde{\slashed{p}}\slashed{A}_{-p}$ in $M_{11}$, $M_{22}$, which, after the same treatment as before, gives us exactly the same actions as Eq.(66),(68)-(70) but with different constants $C_2$.
\begin{equation}\label{eq42}
  C_2=i\int_{\Lambda}\frac{d^4k}{(2\pi)^4}[\hat{f_{k,0}}(+)]k_{4}.
\end{equation}
Taking into account the correction of the constants, we finally obtain all the topological nontrivial actions in the second order calculation.
\begin{eqnarray}
  S^{\prime}_{eff}[+,+] &=& -C\int d^4x\epsilon^{ijk}A_{i}(x)\partial_jA_k(x), \\
  S^{\prime}_{eff}[+,-] &=& C\int d^4x\epsilon^{ijk}A_{i}(x)\partial_jA_k(x), \\
  S^{\prime}_{eff}[-,+] &=& C\int d^4x\epsilon^{ijk}A_{i}(x)\partial_jA_k(x), \\
  S^{\prime}_{eff}[-,-] &=& -C\int d^4x\epsilon^{ijk}A_{i}(x)\partial_jA_k(x),
\end{eqnarray}
where $C=mC_1+C_2$ \cite{rui}.
As has been demonstrated, the above actions lead to bulk currents that are exactly cancelled by each other. However, these Chern-Simons terms become nontrivial and can generate surface states if one considers a system with finite boundary. Two interesting conclusions can be expected. First, since the $\mathrm{U}(1)$ gauge field introduced here does not couple to electrons but to the Bogoliubov quasi-particles, all the excitations in the induced surface states will be Bogoliubov quasi-particles, i.e., the surface states indicated by Eq.(72)-(75) are actually SABS. Second, the zero energy excitation in the SABS is an equal-weight sum of an electron spinor and a hole spinor \cite{Zhou}, leading to the Majorana excitation. Third, due to the opposite sign between $S^{\prime}_{eff}[+,+]$ and $S^{\prime}_{eff}[-,+]$ (or $S^{\prime}_{eff}[+,-]$ and $S^{\prime}_{eff}[-,-]$), we expect the left-handed SABS and the right-handed SABS should enjoy an opposite slope in their energy spectrum. The above expectation purely comes from the field theoretical analysis. In later section, we will confirm these conclusions by studying a lattice model

The above results revealed in the SWM state has a close counterpart in the quantum electrodynamics (QED) \cite{Redlicha,Redlichb,Peskill}. In that case, the renormalization of the effective QED action at zero fermion mass (due to the ultraviolet divergence) gives us an induced Chern-Simons term in the Pauli-Villars term. The Pauli-Villars mass breaks the parity symmetry, leading to the parity anomaly. Here the SC term brings us a mass, which leads to a similar induced Chern-Simons term in the dressed action. However, different from the QED case, SWM state enjoys both left-handed and right-handed sectors due to the original two Weyl points (with opposite chirality). In this case, even though the Chern-Simon term in the left or right sector breaks the parity symmetry, the two Chern-Simons terms $S^{\prime}_{eff}[+,+]$ and $S^{\prime}_{eff}[-,+]$ (or $S^{\prime}_{eff}[+,-]$ and $S^{\prime}_{eff}[-,-]$ ) are parity symmetric to each other, restoring the parity symmetry of the whole SWM system.

\section{Relation between Quantum anomalies and surface states}
\subsection{The topological surface action}
Chern-Simons theory of the external gauge field  usually indicates the emergence of a metallic surface state of the studied system. For example, the axion dynamics and the Chern-Simons action of the 3D topological insulators is responsible for the surface Dirac cone \cite{Qi}, the $3+1$D Chern-Simons term of the WSM phase suggests the Fermi arc in the material surface \cite{Zyuzin}. Here, we find that the SWM phase enjoys two types of nontrivial Chern-Simons actions Eq.\eqref{eq26} and Eq.(72)-(75), originated from two different quantum anomalies. Therefore, we expect the emergence of robust surface states in the SWM state.

To verify to expected surface states, we need to first demonstrate the bulk-surface correspondence in the SWM, i.e., to construct a route, following which, one can obtain a surface state action from the bulk action. In previous sections, we introduced a $\mathrm{U}(1)$ gauge field to the bulk states and then integrated out the matter fields, leaving us a gauge field theory that reflects the bulk property of the SWM. Now let us assume a boundary at $x=0$ (y-z plane), so that the total action of this semi-infinite sized SWM can be written down as
\begin{equation}\label{eqn2}
\begin{split}
  S_{sm}&=\int d\tau d\mathbf{r} \Phi^{\dagger}_{r}[\partial_{\tau}+\tau^z\sigma^z(-i\partial_z)+\sigma^zQ_z(x)\\
  &+s^x\tau^{+}\sigma^z\Delta(x) e^{-2iQ_zz}+s^x\tau^{-}\sigma^z\Delta(x) e^{2iQ_zz}\\
  &+s^z\tau^z\sigma\cdot(-i\nabla_{\parallel})-\mu(x) s^z]\Phi_{r},
\end{split}
\end{equation}
where $Q_z(x)=Q_z\kappa(x)$, $\mu(x)=\mu\kappa(x)$ and $\Delta(x)=\Delta\kappa(x)$, with $\kappa(x)$ being the Heaviside step function. For $x<0$, the action describes a gapless Dirac fermion, which is topologically equivalent to the vacuum \cite{Zyuzin,Goswami}. For $x>0$, we have the SWM state. Now all the above calculations can be performed based on $S_{sm}$. Since $\delta S$ is proportional to $Q_z$, for $x<0$, $Q_z=0$ leads to a vanishing action. Also, for $x<0$ we have $m=0$ so that integral in Eq.\eqref{eq42} leads to $C_2=0$, that further generates $C=mC_1+C_2=0$. Since $S^{\prime}_{eff}[\pm,\pm]$ are proportional to $C$, all the topological actions become zero in the region $x<0$.  After the above consideration, the chiral-anomaly-induced action of the external gauge field with a boundary, reads,
\begin{equation}\label{eqn3}
  \delta S=\frac{e^2}{4\pi^2}\int d^4x\epsilon^{3\nu\alpha\beta}Q_z\kappa(x)A_{\nu}\partial_{\alpha}A_{\beta},
\end{equation}
where a partial integral has been performed. Additionally, the topological actions due to the parity-like anomaly become $S^{\prime}_{eff}[+,+]=-S^{\prime}_{eff}[+,-]=-S^{\prime}_{eff}[-,+]=S^{\prime}_{eff}[-,-]$, with
\begin{equation}\label{eqn4}
  S^{\prime}_{eff}[+,+]=-C\int d^4x\epsilon^{ijk}\kappa(x) A_{i}(x)\partial_{j}A_{k}(x),
\end{equation}

We know that in a uniform infinite-sized medium, the gauge field theory should be gauge invariant, i.e., invariant under transformation $A_{\mu}\rightarrow A_{\mu}+\partial_{\mu}f(x)$, with $f(x)$ being a generic scalar function. However, in the presence of the boundary, the gauge symmetry is broken. Inserting the gauge transformation into Eq.\eqref{eqn3}, and performing the partial integral, an additional term associated to the boundary will emerge, i.e.,
\begin{equation}\label{eqn5}
\delta S_{bd}=-\frac{e^2Q_z}{2\pi^2}\int d^4x\epsilon^{31\alpha\beta}f(x)\delta(x_1)\partial_{\alpha}A_{\beta},
\end{equation}
where the delta function $\delta(x_1)$ comes from the partial derivation, $\partial_{\nu}\kappa(x_1)=\delta_{\nu,1}\delta(x_1)$, and it constraints the action to the boundary surface. Similar action describing the gauge field property in the boundary is found and investigated in the WSM phase by Ref.\cite{Goswami}, where it is shown that this term is exactly canceled by the chiral anomaly from the surface states in the boundary due to the Callan-Harvey mechanism \cite{xgwen,callan}. Due to this reason, Eq.\eqref{eqn5} clearly shows the occurrence of a surface state. Moreover, more information can be shown by the surface current, $j^{\beta}=\delta(\delta S_{bd})/\delta A_{\beta}$.
\begin{equation}\label{eqn6}
  j^{\beta}=\frac{e^2Q_z}{2\pi^2}\epsilon^{31\alpha\beta}\partial_{\alpha}f(x)|_{x_1\rightarrow0}.
\end{equation}
This current is not proportional to any external field, so that it reflects the intrinsic topological property of the SWM state. Besides, the sign in front of $j^{\beta}$ is the reflection of the energy dispersion slope (chirality) of the surface state. As is clear from the above equation, $j^3=0$ means that the dispersion slope is zero along $z$ direction, while $j^2\neq0$ suggests that the surface state is dispersive along $y$ axis. These are the main properties of the Fermi arc previously found in WSM phase \cite{Wan}. Here, we have proved their existence in the SWMs.

What we are more interested is the surface action due to the parity-like anomaly. Similarly, inserting the gauge transformation into Eq.\eqref{eqn4} and completing the partial integral, we obtain $S_{bd}[+,+]=-S_{bd}[+,-]=-S_{bd}[-,+]=S_{bd}[-,-]$, with
\begin{equation}\label{eqn7}
  S_{bd}[+,+]=C\int d^4x\epsilon^{1jk}f(x)\delta(x_1)\partial_{j}A_{k},
\end{equation}
Similar to $\delta S_{bd}$, $S_{bd}[\pm,\pm]$ also convincingly justify the occurrence of four more surface states in the SWM boundary. The currents can be arrived at after performing the variation, leading to $j^{k}[+,+]=-j^k[+,-]=-j^k[-,+]=j^k[-,-]$, with
\begin{equation}\label{eqn8}
  j^{k}[+,+]=-C\epsilon^{1jk}\partial_{j}f(x)|_{x_1\rightarrow0}.
\end{equation}
Since $S^{\prime}_{eff}[+,-]$ and $S^{\prime}_{eff}[--]$ are particle-hole symmetric to $S^{\prime}_{eff}[+,+]$ and $S^{\prime}_{eff}[-,+]$, respectively. It is sufficient to discuss only the current $j^k[+,+]$ and $j^k[-,+]$, which describes the left- and right-handed surface currents, respectively. It is clear that they deviate from each other by a minus sign. This means the surface states would enjoy an opposite chirality in their dispersion spectrum, i.e., if one is left-moving then the other must be right-moving. We also note that there is a major difference between the two currents in Eq.\eqref{eqn6} and Eq.\eqref{eqn8}. As we know from the calculation in the last section, the gauge field $A_{\mu}$ in Eq. \eqref{eqn5} is coupled to the electron fields or hole fields in the Grassmann field $\Phi_{r}$, while the gauge field $A_{\mu}$ in Eq.\eqref{eqn7} is coupled to the Bogoliubov quasi-particles in the Grassmann field $\Phi^{\prime\prime}_{r}$. This fact is important since it leads to different physical meanings between the two types of surface states indicated by Eq.\eqref{eqn6} and Eq.\eqref{eqn8} respectively, i.e., the excitations in the first type of surface state are electrons while that in the second type are Bogoliubov quasi-particles. This is in agreement with our expectation because the first surface state (from Eq.\eqref{eqn6}) is actually the Fermi arc inherited from the original WSM phase. The electrons in the Fermi arc do not form pairs and thus keep their electron properties. In contrary, the second type of surface state (from Eq.\eqref{eqn8}) reside in the SC gap and is the result of the pairing instability, they are in essence the SABSs. Remarkably, the $E=0$ quasi-particle excitations in the SABSs are the Majorana fermions, since they are equal-weight sum of the electron spinor and the hole spinor \cite{Zhou,Bednik}.
\subsection{Verification by the tight-binding model}
The above field theoretical results can be verified by analyzing the tight-binding model, from which one can also numerically obtain the surface states. We now utilize the following lattice model to describe the normal state band structure \cite{Cho}
\begin{eqnarray}
H_{0}(\mathbf{k})&=&t(\sigma^{x}\sin k_{x}+\sigma^{y}\sin k_{y})+t_{z}(\cos k_{z}-\cos Q)\sigma^z   \notag \\
&&+m(2-\cos k_{x}-\cos k_{y})\sigma^z-\mu\sigma^0.
\end{eqnarray}
The basis is taken as $\{c^{\dagger}_{\uparrow}(\mathbf{k}),c^{\dagger}_{\downarrow}(\mathbf{k})\}$.
The model can be written compactly as
\begin{equation}
H_{0}(\mathbf{k})=\sum\limits_{\alpha=0}^{3}d_{\alpha}(\mathbf{k})\sigma^{\alpha},
\end{equation}
where $d_{0}(\mathbf{k})=\mu$, $d_{1}(\mathbf{k})=t\sin k_{x}$, $d_{2}(\mathbf{k})=t\sin k_{y}$, and $d_{3}(\mathbf{k})=t_{z}(\cos k_{z}-\cos Q)+m(2-\cos k_{x}-\cos k_{y})$.
For large $m$, only one pair of Weyl points exist in the BZ, at $\mathbf{P}_{\pm}=(0,0,\pm Q)$. For small $\mu$, the Fermi surface consists of two disconnected spherical pockets centered at $\mathbf{P}_{+}$ and $\mathbf{P}_{-}$, respectively. Without losing any generality, we set $\mu>0$ and focus on low energy properties related to the upper band with dispersion
\begin{equation}
E_{+}(\mathbf{k})=-\mu+\sqrt{d_{1}^{2}(\mathbf{k})+d_{2}^{2}(\mathbf{k})+d_{3}^{2}(\mathbf{k})}.
\end{equation}
We represent the creation operator of this band as $\tilde{c}^{\dagger}_{+}(\mathbf{k})$. Introducing the relative wave vectors $\mathbf{q}=\mathbf{k}-\mathbf{P}_{\pm}$ for states close to either of the two Weyl points, the FFLO pairing order parameter can be written as\cite{Cho}
\begin{equation}
\Delta_{\pm}c^{\dagger}_{\alpha}(\mathbf{q}+\mathbf{P}_{\pm})(i\sigma^{y})^{\alpha\beta}c^{\dagger}_{\beta}(-\mathbf{q}+\mathbf{P}_{\pm}),
\end{equation}
where $\Delta_{+}$ and $\Delta_{-}$ are the pairing amplitudes of the FFLO state in the two Fermi pockets. Projecting the above pairing term to the two spherical pockets of the Fermi surface around $\mathbf{P}_{+}$ and $\mathbf{P}_{-}$, we have the following effective pairing terms
\begin{equation}\label{eqh1}
\Delta_{\nu}\tilde{c}^{\dagger}_{+}(\mathbf{q}+\mathbf{P}_{\nu})\text{sgn}(t)\frac{q_{x}+iq_{y}}{\sqrt{q^{2}_{x}+q^{2}_{y}}}\tilde{c}^{\dagger}_{+}(-\mathbf{q}+\mathbf{P}_{\nu}),
\end{equation}
where $\nu=\pm$ labels the two Weyl points and the two surrounding Fermi pockets, and $\text{sgn}(t)$ is the sign of $t$. For any wave vector on the Fermi pocket centered at $\mathbf{P}_{\nu}$, the absolute value of the paring amplitude is a constant, $|\Delta_{\nu}|$. So, this state is fully gapped, if we introduce pairing in the neighborhood of the two Fermi pockets. More importantly, from well-known results for the $p+ip$ pairing, we know that there are two pairs of SABSs on a surface parallel to the $z$ axis\cite{read00}, one is around $\mathbf{P}_{+}$ and the other is around $\mathbf{P}_{+}$. This conclusion is in agreement with the results from the parity-like anomaly, i.e., Eq.\eqref{eqn8}.
\begin{figure}[tbp]
\includegraphics[width=3.4in]{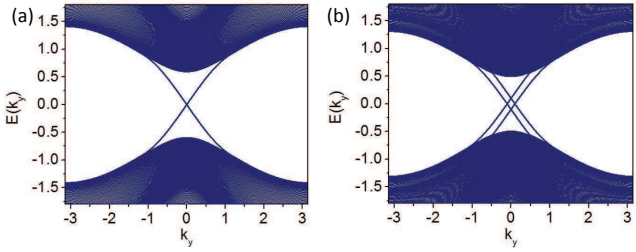}
\caption{(color online). Numerical results of the energy spectrum along $k_y$ ($k_z=0$) of the tight-binding model for different chemical potential. A thin film with 200 layers along the x direction is considered. The parameters chosen are $t=-1$, $t_z=-2$, $Q=\pi/4$, $m=1$ and $\Delta_{+}=\Delta_{-}=0.1$. The chemical potential is $\mu=0$ in (a) figure while $\mu=0.1$ in (b) figure.}
\end{figure}

Moreover, one can obtain the surface modes from the above tight-binding model. Ref.\cite{Zhou} has calculated the eigenvalues of the above Hamiltonian with two surface boundaries normal to the $x$ direction (for $\mu\neq0$). The dispersion along $k_z$ is shown in the Fig.5(a) in Ref.\cite{Zhou}. It is found that four chiral surface states show up in the surface boundary, i.e.,
two occur in the SC gap around the left WP, while two other surface states occur in the SC gap around the right WP. In both the SC gaps, one surface state is particle-hole symmetric and the other is particle-hole antisymmetric. These states are the Andreev surface states, and they are composite states of electron and holes. Moreover, let us focus on the particle-hole symmetric ones, as can be found in Fig.5(a) in Ref.\cite{Zhou}, the surface states near the left-handed and right-handed WP enjoy the opposite Fermi velocity: one is left-moving and the other is right-moving.  Recalling that in the topological field theory, we have derived four branches of Chern-Simons actions in the SC gaps, \textit{i.e.}, Eq.(72)-(75), and they further generates four surface currents $j^k[\pm,\pm]$. As has been pointed out, the currents from the left- and right-handed sectors have the same absolute value but differ by a minus sign, leading to the opposite direction of current response, which manifests itself in the opposite Fermi velocity of the SABS. Hence, we have found out the exact correspondence between the numerical surface states, the low-energy model derived from the tight-binding Hamiltonian, Eq.\eqref{eqh1}, and the topological field theory.

More interestingly, our topological field theory also includes the chiral anomaly, leading to the boundary action $\delta S_{bd}$ in Eq.\eqref{eqn5}. This means that we must be able to find another type of surface state which is the Fermi arc inherited from the WSM state. Since Ref.\cite{Zhou} does not show any signature of this Fermi arc, we perform the numerical diagonalization of the lattice model. In order to show the arc clearly, we intentionally set $\mu=0$ and plot the energy spectrum for a thin film with two hundred layers stacked along x the direction. Fig.3(a) shows the result along $k_y$ direction with $k_z$ fixed to be 0. Clearly, we observe two chiral surface states, where the one with the positive slope and the negative slope  belongs to the upper and lower boundary, respectively. Since the surface state in Fig.3 occurs at $k_z=0$, and the pairing gap only dominates in the region around the two Fermi pockets located at $k_z=\pm Q_z$, the arc observed here cannot be the SABS, which are located in the SC gap. Moreover, if one plots along $k_z$ direction, one will find the dispersion becomes flat, which is the main characteristic of the Fermi arc in WSM state. Hence, this state obtained here can only be identified as the Fermi arc inherited from the original WSM phase.  Moreover, since the chemical potential is $\mu=0$, the original Weyl points are completely gapped out and the bulk SWM phase has a full gap in its energy spectrum. However, the Fermi arc still persists. This numerical observation is consistent with our field theoretical conclusion, where we show that the chiral anomaly does not rely on the Weyl points, and the topological response remains in the SWM state even though the Weyl nodes are gapped out. Moreover, in Fig.3(b), we plot the energy spectrum for $\mu=0.1$, where two Fermi arc surface states belonging to two boundaries are still found. Due to the particle-hole symmetry, two more Fermi arcs emerge, which are the redundancy due to the introduction of Nambu space.

Now we can conclude that, despite being a superconductor in the bulk, our model of the SWM state is topological in two senses. The first is that the original Berry curvature around the Weyl points persists after the onset of superconductivity, so that the Fermi arc state remains. The second is that the superconductivity, by itself, is able to induce four surface Andreev bound states in the SWM boundary. All the results are explained by our topological field theory of the SWM state, which, to the best of our knowledge, is the first field theory of the superconducting Weyl materials.
\section{Quantum anomalies in BCS state}
In the above section, we have discussed the quantum anomalies in the intranode FFLO state in detail. Now we are going to study the simplest internode pairing state, \textit{i.e.}, the s-wave BCS pairing state, whose effective description is given by Eq.\eqref{eq7}. All the method is the same as that used for studying the FFLO case, therefore we only briefly present the results and the discussions in this section.

Obviously, the BCS pairing term $s^x\tau^z\Delta$ is invariant under the local chiral transformation, $\Phi_{\mathbf{r}}\rightarrow e^{-is^0\tau^z\sigma^0\theta(\tau,\mathbf{r})/2}\Phi_{\mathbf{r}}$. The transformed action reads,
\begin{equation}\label{eqt1}
\begin{split}
  S^{\prime}_2&=\sum_{i\omega_n}\int\frac{d^3 k}{(2\pi)^3}\Phi^{\dagger}(i\omega_n,\mathbf{k})
  (-i\omega_n+\tau^z\sigma^zk_z-\mu s^z\\
  &+s^z\tau^z\sigma\cdot\mathbf{k}_{\parallel}+\Delta s^x\sigma^z)\Phi(i\omega_n,\mathbf{k}),
\end{split}
\end{equation}
where we have insert $\theta(\tau,\mathbf{r})=2Q_zz$. If we view the above action as the composition of the superconductor term $s^x\tau^z\Delta$ and the non-superconducting Weyl metal term, two conclusions are obvious. First, the chiral transformation shifts the two Weyl points, and they finally meet each other at $\Gamma$ point, forming a Dirac point with spin-momentum locking. Second, the superconductivity term is invariant. Since the transformed phase has a Dirac cone at $\Gamma$ point and the pairing electrons now come from the electron states of the 3D Dirac cone with momentum $k$ and $-k$, the resulting state can be shown to enjoy a SC gap at $\Gamma$ point. Now, it is clear that, similar to the FFLO SWM state, there are two important questions needed to be addressed. One is the chiral anomaly from the Jacobian of the functional integral measure of the chiral transformation, and the other is the possible parity-like anomaly of the gapped transformed state.

To study the first one, we introduce the Dirac matrices $\gamma^{\mu}$ as before, then the SC term is written into
\begin{equation}\label{eqt2}
  S_{SC2}=-i\int d\tau d\mathbf{r}\overline{\Phi}_{r}\Delta s^x\gamma^5\gamma^4\gamma^3\Phi_{r}.
\end{equation}
Then following the same method as used in the FFLO case, \textit{i.e.}, introducing external gauge field and then evaluating the Jacobian of the functional integral measure of the chiral transformation, one will encounter the trace of Dirac matrices. It is found that, the SC term Eq.\eqref{eqt2} does not have any contribution because the Dirac matrices in front of this term ($\gamma^5\gamma^3\gamma^3$) will make the trace zero. This means that the Jacobian only comes from the nonsuperconducting terms $S^{\prime}_2-S_{SC2}$. Since, in the FFLO state, the anomaly also originates from the non-superconducting Weyl metal term, it then becomes clear that the internode BCS state enjoys exactly the same Jacobian and thus the same effective gauge field action as Eq.\eqref{eq26}. Due to the discussion in the above section, we know that a Fermi arc must be present in the BCS pairing SWM phase, which is due to the chiral anomaly. 

Now we discuss the parity-like anomaly in the BCS state. Using the same method as before, we first perform the canonical transformation $s^{\pm}\rightarrow\sigma^z s^{\pm}$, $\sigma^{\pm}\rightarrow s^z\sigma^{\pm}$, and then introduce the Dirac matrix $\eta$. After diagonalization in the Nambu space, we find that the effective action around the $\Gamma$ points of the BCS type SWM state (after the chiral-transformation) can be reduced exactly to the same form as that in Eq.\eqref{eq30}. The parity-like anomaly, which is responsible for the occurrence of the Andreev surface state, is only dependent on the transformed action. Therefore, in terms of the surface state, we have shown in a general field theoretical way that, the BCS internode pairing SWM phase has the same property as the FFLO intranode pairing SWM phase, \textit{i.e.}, it also enjoys the topological Chern-Simons action, Eq.\eqref{eq26} and Eq.(72)-(75). This accounts for the similar surface states between the FFLO and the BCS states \cite{Bednik}. For the BCS state, more details will be presented by our future work. In the present work, we only discuss the simplest s-wave singlet pairing state, while other pairing states, as raised by Ref.\cite{Cho}, can also be investigated in the same method with a straightforward generalization.
\section{Conclusions}
We have performed a field theoretical investigation of the SWM. Two quantum anomalies are extracted in this state. The first is the chiral anomaly. In the presence of a boundary, the chiral anomaly shows a bulk Chern-Simons action, which further induces a chiral surface state due to the Callan-Harvey mechanism \cite{xgwen,callan}. The second quantum anomaly is due to the onset of superconductivity. It generates four Chern-Simons actions, $S^{\prime}_{eff}[\pm,\pm]$. This anomaly, similar to the parity anomaly in the QED, gives rise to four SABSs, when a finite size system is considered. All the above results are consistent with our tight-binding model calculation. Even though we only discuss the s-wave pairing case, straight forward extension can be made to study other pairing states. Moreover, our method can also be applied on more realistic SWM models, and even the superconducting type \uppercase\expandafter{\romannumeral2} Weyl metals \cite{Soluyanov}, which will be an interesting investigation in the future.
\begin{acknowledgments}
Rui Wang wishes to thank W. P. Su and Hai Li for fruitful discussion. L. Hao acknowledges T. Zhou for useful communication. This work is supported by the Texas Center for Superconductivity at the University of Houston and the Robert A. Welch Foundation (Grant No. E-1146) and the National Natural Science Foundation of China (Grants No. 60825402, No. 11574217 and No. 11204035).
\end{acknowledgments}




\begin{thebibliography}{99}
\bibitem{ruisec} For $\Delta\neq0$, a full gap is opened. This is the necessary condition required for the following calculations. As long as we are in the full gapped phase, the bulk topology is the same. Therefore, all the following results are applicable for larger $\Delta$.
\bibitem{rui} Here, it is not necessary to calculate the specific value of the constant $C$ because, as will be clear later, the physical manifestation of the above actions is the Majorana mode located at the surface boundary, which does not have well-defined charge
\end{thebibliography}
\end{document}